\documentclass[runningheads]{llncs}
\usepackage{graphicx} 
\usepackage{booktabs}
\usepackage{xspace}
\usepackage{arydshln}
\usepackage{hyperref}
\usepackage{makecell}
\usepackage{multirow}
\usepackage{amssymb}
\usepackage{xcolor}
\usepackage{times}
\usepackage{amsmath}
\usepackage{soul}

\newcommand{\partitle}[1]{\vspace{2mm}\noindent\textbf{#1}}
\newcommand{\modelname}{\texttt{UniFed}\xspace}
\newcommand{\ie}{\emph{i.e.,~}}
\newcommand{\eg}{\emph{e.g.,~}}

\begin{document}
%
\title{\modelname: A Universal Federation of a Mixture of Highly Heterogeneous Medical Image Classification Tasks}
\titlerunning{\modelname: A Universal Federation of a Mixture of Heterogeneous Images}
%
\author{Atefe Hassani \and Islem Rekik}
%
\institute{BASIRA Lab, Imperial-X and Department of Computing, Imperial College London, \\ London, UK \\
\email{hasaniatefe0@gmail.com,i.rekik@imperial.ac.uk}\\
}

\maketitle              
\begin{abstract}
A fundamental challenge in federated learning lies in mixing heterogeneous datasets and classification tasks while minimizing the high communication cost caused by clients as well as the exchange of weight updates with the server over a fixed number of rounds. This results in divergent model convergence rates and performance, which may hinder their deployment in precision medicine. In real-world scenarios, client data is collected from different hospitals with extremely varying components (e.g., imaging modality, organ type, etc). Previous studies often overlooked the convoluted heterogeneity during the training stage where the target learning tasks vary across clients as well as the dataset type and their distributions. To address such limitations, we \emph{unprecedentedly} introduce \modelname, a universal federated learning paradigm that aims to classify any disease from any imaging modality. \modelname also handles the issue of varying convergence times in the client-specific optimization based on the complexity of their learning tasks. Specifically, by dynamically adjusting both local and global models, \modelname considers the varying task complexities of clients and the server, enhancing its adaptability to real-world scenarios, thereby mitigating issues related to overtraining and excessive communication. Furthermore, our framework incorporates a sequential model transfer mechanism that takes into account the diverse tasks among hospitals and a dynamic task-complexity based ordering. We demonstrate the superiority of our framework in terms of accuracy, communication cost, and convergence time over relevant benchmarks in diagnosing retina, histopathology, and liver tumour diseases under federated learning. Our \modelname code is available at \url{https://github.com/basiralab/UniFed}.


\end{abstract}
\section{Introduction}
\label{sec:intro}
Federated learning (FL) is a powerful decentralized framework that enables clients to collaboratively train a global model without sharing their local datasets, thereby preserving privacy and security \cite{mcmahan2017communication,ghilea2023replica}. This approach reduces communication costs and enhances user privacy by eliminating the need to transmit raw data to a central server \cite{mcmahan2017communication,ezzeldin2023fairfed}.
In contrast to traditional centralized learning approaches, FL processes data locally, directly where it is generated, such as within distributed hospital systems. While early FL research primarily focused on single-task learning \cite{zhang2023fedala,isik2022sparse,mcmahan2017communication}, recent investigations have expanded into federated multi-task learning (FMTL).

FMTL methods have recently emerged as an alternative approach to learning personalized models in the federated setting to solve multiple related learning tasks while exploiting patterns across tasks (\eg \cite{marfoq2021federated} defines federated MTL as a penalized optimization problem, where the penalization term models relationships among tasks (clients)) \cite{smith2017federated}. While existing FMTL approaches primarily address statistical challenges \cite{marfoq2021federated,smith2017federated}, recent studies \cite{cai2023many,chen2023fedbone,zhuang2023mas} have highlighted the importance of \emph{task heterogeneity}, particularly for dense predictions such as semantic segmentation and depth estimation \cite{chen2018encoder,wang2021pyramid}. It also has been broadly used in various domains such as healthcare \cite{baytas2016asynchronous,suresh2018learning}, financial forecasting \cite{cheng2020federated}, and IoT computing \cite{mansour2020three}.

Despite the successful FL applications, we still face critical challenges. \emph{First}, multitasking in a single hospital presents significant problems where each hospital maintains a task-specific model for each task. Therefore, directly deploying multiple models for each task in hospitals is impractical. \emph{Second}, the substantial communication cost of exchanging weight updates from clients to the server in each round results in divergent model convergence rates and performance. \emph{Third}, \emph{data and task heterogeneity} pose significant challenges in FL. Clients naturally have non-identical independent (Non-IID) data, leading to imbalanced characteristics and unequal training contributions. Furthermore, when multiple tasks are distributed across clients without specific preferences, the learning divergence between clients is further intensified. Consequently, the update rate should vary based on the difficulty of the task assigned to each client. Several groundbreaking studies have made efforts to mitigate these challenges through various solutions. Jia et al.~\cite{jia2024fedlps} proposed FedLPS, a method that reduces training costs on edge devices with multiple tasks by splitting local models into shared encoders and task-specific predictors. Niu et al.~\cite{niu2023flrce} applied globally (communication round) early stopping by selecting clients with more significant effects, the global model can converge in fewer rounds. FLrce \cite{niu2023flrce} reduced the overall resource computation resources by decreasing the total training rounds, thereby limiting the overall resource usage.
However, FedLPS and FLrce, do not involve dynamic updates of clients considering their varying contributions, making it difficult to save storage resources or communication overhead. They only focus on reducing global communication overhead between clients and the server.

To address these critical challenges, we propose \modelname, a dynamic sequential federation for universal medical image classification (i.e., classify any imaging modality of any disease), exploiting the client convergence rate and its task difficulty.
Our approach aims to prevent overtraining and undertraining while reducing resource usage in multi-task scenarios.
This form of MTL considers each dataset as a separate and non-related classification task where each client is trained on \emph{only} a particular task while aiming to generalize to other unseen tasks. 
The main contributions of this work are:

\begin{figure}[!th]
  \centering
   \includegraphics[scale=0.44]{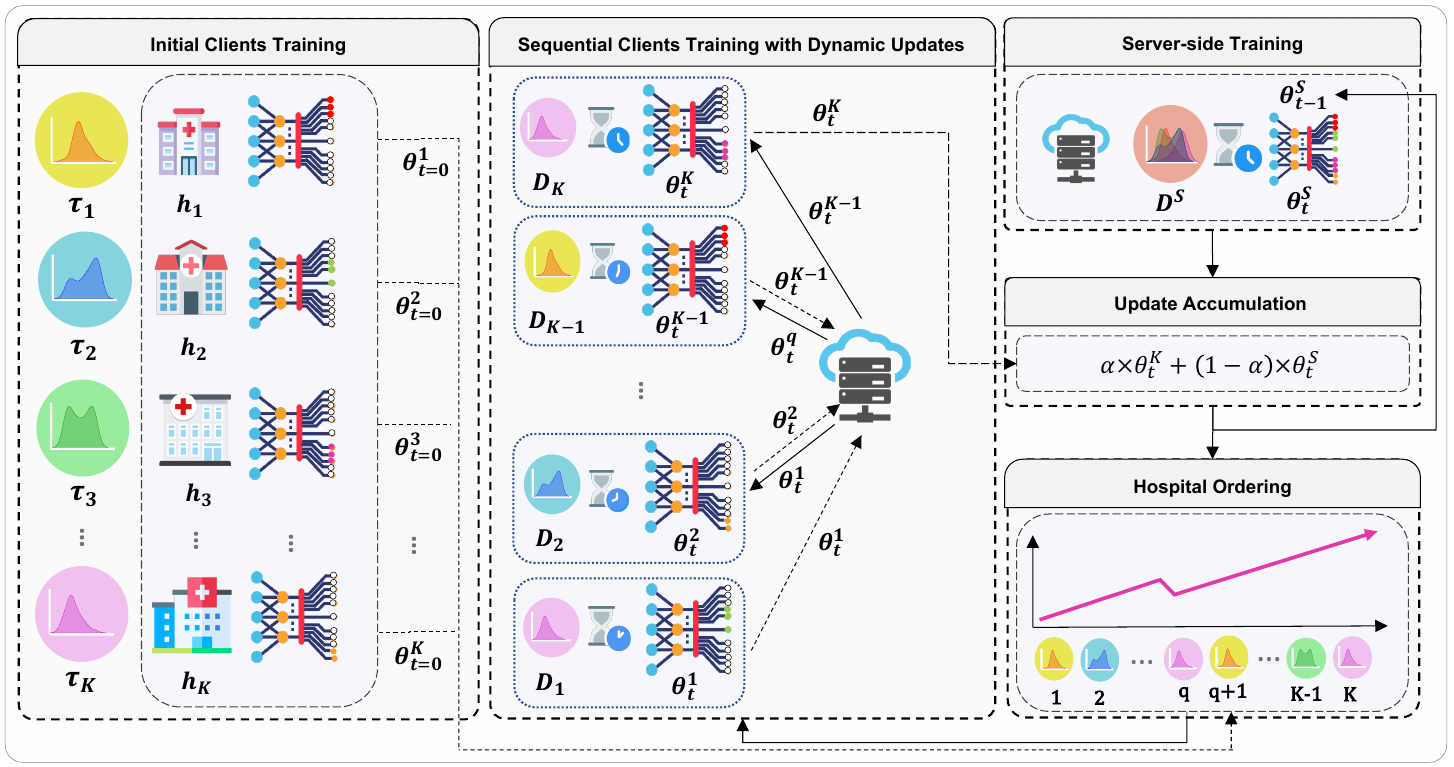}
  \caption{An overview of \modelname framework. Each hospital receives the initial model from the server and trains it for one epoch only in the first round. Given the initially ordered clients, we dynamically update each hospital by sequential model passing based on task complexity. The server is first trained on an independent public mixed dataset, then receives the last hospital parameters $\theta_t^K$ and regularizes it by taking the average of server parameters $\theta_t^S$.}
\label{fig:fram}
\end{figure}

\begin{enumerate}
    \item We propose a novel FL framework, \modelname, \textbf{learning from a mixture of highly heterogeneous tasks} by introducing a \emph{loss-guided dynamic and sequential} model exchange between the server and clients with application to medical image classification.
    \item We further remedy the variance in convergence time across clients during our sequential exchange process while being agnostic to task complexity.  
    \item Our framework guarantees efficient communication from clients to the server without sacrificing accuracy.
    \item We show the outperformance of \modelname on various medical imaging datasets including  OCTMNIST, OrganAMNIST, and TissueMNIST datasets with classification tasks \emph{under both strongly and moderately Non-IID data splits}; and show improvement in communication cost and convergence time over relevant baselines.
\end{enumerate}

\partitle{Related Work.}
\label{sec:relatedworkd}
FL involves collaboratively training a global model over distributed local data among clients while keeping the raw data private and local to the participating client. FL commonly faced the heterogeneity issue where the $i^{th}$ client's data is different from the $j^{th}$ client's data in several aspects including statistics, systems, data domains, and models \cite{gao2022survey}. Our paper focuses on two heterogeneous aspects: statistical heterogeneity and system heterogeneity. Statistical heterogeneity causes data to be distributed Non-IID among edge devices, which raises the difficulties of model convergence \cite{hsieh2020non}, while system heterogeneity results in differences in the capabilities of these devices. 

In recent years, FMTL has emerged as a promising approach and has attracted much attention to address the challenges of federated learning in scenarios where multiple related tasks need to be learned simultaneously. They often encounter challenges such as high communication costs and privacy concerns, which prevent data transfer to a central node. FMTL proposes to train separate models for each client while sharing knowledge among them, focusing on the privacy of clients \cite{smith2017federated}. \cite{marfoq2021federated} proposed federated multi-task learning under the assumption that each local data distribution is a mixture of some underlying distribution. They showed the advantage of distilling the client's knowledge to each client. FedBone \cite{chen2023fedbone} utilizes split learning for large-scale federated training on edge clients and adapting to heterogeneous tasks and introduces GPAggregation method to enhance the global model generalization. FedLPS \cite{jia2024fedlps} divided the local model into a shareable encoder and task-specific predictors, drawing from transfer learning principles. It also incorporates a channel-wise model pruning algorithm to reduce the compute overhead while considering data and system heterogeneity. Furthermore, FedLPS introduces a novel heterogeneous model aggregation algorithm to combine the diverse predictors effectively.

However, existing works only explored reducing resource consumption and enhancing the global model generalization for multiple tasks on edge devices in FL without considering task complexity and dynamic updates of clients and the global model. In this paper, we address this gap and propose dynamic updates and training of clients by ordering them based on task difficulty to reduce communication costs and enhance the local hospital model generalizability to unseen diagnostic tasks among others listed in our contributions above.

\section{Method}
\label{sec:method}
Fig.~\ref{fig:fram} provides an overview of the proposed \modelname framework. We design \modelname \textbf{(Supp.~Algorithm 1)} to efficiently train a mixture of highly heterogeneous medical image classification tasks in the context of FL. \modelname involves an effective technique to schedule the sequential training of clients, avoiding the reliance on random order selection \cite{chang2018distributed,li2024convergence}.
Furthermore, we propose a local and global dynamic update technique that mitigates overtraining and high communication costs between clients and server.

\partitle{Problem Definition.} Let $ \mathcal{D} = \{D_k\}_{k=1}^K $ 
denote a set of $K$ datasets, where each has a pre-assigned classification task $ \tau_k$ . Each dataset $D_k = \{ (\mathbf{x}^{(k)}_i,y^{(k)}_i) \}_{i=1}^{n_k}$ contains $n_k$ samples, where $\mathbf{x}^{(k)}_i$ denotes the input features, and $y^{(k)}_i$ the corresponding output label. The data is non-independent identically distributed (Non-IID). Each hospital is trained on a single classification task but aims to generalize its model to other unseen tasks. For instance, while one task aims to detect retinal diseases using optical coherence tomography (OCT) images, another task aims to classify liver tumour disease using computed tomography (CT) images.

We denote by $\theta_t^k$ the model weights of hospital $h_k$ which is trained on the single-task dataset $D_k$ at round $t$. The objective is to solve the following optimization problem and obtain reasonable local models which can be formalized for a given loss function $l(.)$ as:

\begin{align}
&\min_{\theta\in\mathbb{R}^{d}} \Bigg[ \mathcal{L} (\mathbf{\theta}) := \frac{1}{K}\sum_{k=1}^K \mathcal{L}_k (\theta) \Bigg], \text{where} \, \mathcal{L}_k(\theta) = \sum_{(\mathbf{x},y) \in {D_k}} l(\mathbf{x},y,\theta) 
\end{align}

The final objective is to simultaneously train a global model $\theta^*$ with a low error on all tasks in $\mathcal{D}$, leveraging shared learned representations from distinct but related tasks. The idea of federated learning is to enable each hospital to benefit from data samples available at other hospitals to get a better estimation of $\mathcal{L}(\theta)$, and therefore get a model with a better generalization ability to unseen examples.

\partitle{Hospital Ordering.} 
Prior to the start of the local training of each client, the server broadcasts an initial model with random seed $\theta_{0}$ to all hospitals, where models are trained in parallel across clients \cite{mcmahan2017communication}. Each hospital $h_k$ trains locally for one epoch resulting in $\theta_0^k$ (this training is done only at the first global communication round), these models are then sent back to the server.
Next, the server performs a lifelong learning technique known as Curriculum Learning (CL) \cite{bengio2009curriculum,soviany2022curriculum}, which consists of training a model on tasks ordered by their complexity. Similar to how a person would learn in the real world \cite{elman1993learning,fan2017self}, a model trained with this simple rule will end up learning simpler and more general features before moving on to harder ones. We choose the \emph{slope of the loss} during training as a simplified metric for ordering hospitals based on their task complexity.
Each hospital $h_k$ computes $s_k$ using loss values stored in a vector $L$ indexed by the batch number. We denote by $L_k^b$, hospital $h_k$’s loss value for a batch $b$:
\begin{align}
s_k=\frac{\sum_{b=1}^{\mid{L_k}\mid}(b-\bar{b})(L_k^b - \bar{L_k})}{\sum_{b=1}^{\mid{L_k}\mid}(b-\bar{b})^2}.
\end{align}

In short, $s_{k}$ represents the slope of the first-degree polynomial that best fits the loss curve. A lower loss slope suggests a faster convergence rate whereas a higher one indicates a slow convergence rate. This score reflects the task difficulty under the assumption that the rate of convergence of the model $k$ when trained on $D_{k}$ is indicative enough of dataset $D_{k}$ complexity. By transmitting this score to the central server, the server can dynamically apply CL in each round to optimize the scheduling of the next training phase.

\partitle{Dynamic Update.}
Inspired by \cite{prechelt2002early}, we implement \emph{dynamic training} in FL to mitigate overfitting and underfitting, enhancing convergence, generalization, and consistency across hospitals. Dynamic updates allow for varying convergence times, essential for hospitals with unique classification tasks and no class overlap. Each hospital's training is tailored to its task complexity rather than a fixed number of epochs.

To describe the convergence conditions, we evaluate the per-sample loss $l$ of each hospital (which has been ordered based on the task complexity) on the validation set during consecutive strips, e.g., after each seventh and tenth epoch for local and global training, respectively. 
Next, we stop the training process when the validation loss increases for $z$ consecutive epochs, $l_{val}(e) > l_{val}(e-z)$, or remains nearly unchanged compared to the previous value within $z$ epochs, where $e$ denotes the current epoch.
This approach enables each hospital to prolong its training until meeting the convergence condition depending on its task difficulty and data complexity. Besides, we observe that once we optimize all hospitals till convergence, the convergence time changes in the subsequent hospitals and rounds, as shown in \textbf{Supp.~Table} 1. All convergence criteria end \emph{dynamically} when the client's model converges during training and the result of the training is then a set of weights which shows the lowest validation error.

\partitle{Sequential Clients Training.}
Traditional sequential federated learning (SFL) methods \cite{li2024convergence} suffer from client drift and poor generalization on individual tasks, leading to performance degradation \cite{kang2023one}.
\modelname addresses this by incorporating SFL and CL techniques, dynamically updating hospitals based on task difficulty. After obtaining the ordering of hospitals’ task complexities, the server starts a global communication round, in which, it sends the model from one hospital to another. At a global round $t$, the server $S$ keeps a copy of $\theta_{t-1}$ denoted by $\theta_{t-1}^S$.
Following the ordering obtained in the previous step, $\theta_{t-1}$ is sent to the first hospital $h_{1}$ as $\theta_{t}$. At $h_{1}$, $\theta_{t}$ is dynamically trained on local dataset $D_1$ for $E$ local epochs by applying the aforementioned dynamic update schema. The resulting model $\theta_t^{1}$ is sent to the server where it is kept for performing CL again to \emph{reorder them for the next round}. Next, the server broadcasts  $\theta_t^{1}$ to the next hospital and replaces it with the last trained model. The process repeats until we go over all hospitals, in a sequential manner with dynamic updates and following the dynamic ordering. In round $t$, the obtained final model $\theta_t^{K}$ has been trained sequentially on $D_{K}$, full participation of clients, with a high probability of observing all learning tasks. We thus, effectively and dynamically train a model on a mixture of datasets while preserving privacy and security by only communicating model weights and score values.

\partitle{Server Training and Update Accumulation.} We incorporate server-side learning with a small amount of training data into our framework following \cite{mai2022federated,song2022personalized} and incrementally distil this knowledge into the global model through the federated learning process, without sending any global or shared dataset to the clients.
At each round, following the dynamic sequential training step, the server receives $\theta_t^{K}$ (the final model received by the server). It is regularized by taking the average with $\theta_t^S$, a model that is kept at the server side and is dynamically trained on a small local mixed set $D^{S}$ instead of sharing the data among all the clients \cite{song2022personalized,mai2022federated}, which contains mixed data from all task distributions, $\theta_{t+1}=\alpha \theta_t^{K} + (1-\alpha)\theta_t^S$, where $\alpha$ denotes the mixing weight. By adjusting $\alpha$, we can prioritize either the clients' or the server's regularization term. In our experiments, we set $\alpha$ to 0.7. The accumulated model is then sent to the first client after re-ranking all hospitals, and then dynamically trained based on its task complexity following dynamic update and sequential training.

\section{Results}
\label{sec:experiments}
\partitle{Dataset.}
To simulate our specific scenario, where hospitals have disparate classification tasks and aim to collaboratively train a universal model capable of predicting all tasks within the network, it is essential to conduct experiments using real-world medical data from \emph{distinct modalities}. We simulate image classification tasks using three public medical imaging datasets including OCTMNIST with 4 classes, OrganAMNIST with 11 classes, and TissueMNIST with 8 classes sampled from the MedMNIST2D \cite{medmnistv2} dataset. 
Since different hospitals have different tasks which implies heterogeneous label distribution across tasks, as illustrated in \textbf{Supp.~Fig 2}, we define two variants of Non-IID settings. Specifically, we define the heterogeneous settings using two scenarios: \emph{strongly and moderately Non-IID}.
In the strongly Non-IID setting, each hospital has different portions of labels. On the other hand, we assign the same portion of labels to each hospital in a moderately Non-IID scenario. For each dataset, we randomly sample 5\% of the Non-IID data partitions and reserve them as server data, while using the remaining 95\% partitions as client datasets.
For all hospitals, we split each local dataset into training (70\%), validation (10\%), and test (20\%) sets in a stratified way.

\begin{table}[!h]
\centering
\caption{Comparison of models performance (\%) in both strongly and moderately Non-IID settings.}\label{tab_perf}
\resizebox{\columnwidth}{!}{%
\begin{tabular}{llcccclcccclcccc}
\toprule
\multirow{2}{*}{\makecell{Data\\Partition}} & \multirow{2}{*}{Method}  & \multicolumn{4}{c}{CNN} && \multicolumn{4}{c}{VGG11} && \multicolumn{4}{c}{ResNet18} \\
\cline{3-6}\cline{8-11}\cline{13-16}
&& {Acc} & {F1} & {Sens} & {Spec} && {Acc} & {F1} & {Sens} & {Spec} && {Acc} & {F1} & {Sens} & {Spec} \\
\midrule
& NoFed  & 68.17 & 54.15 & 56.42 & 96.63   &&   77.10 & 65.90 & 67.31 & 97.00   && 74.94 & 61.26 & 63.39 & 96.94 \\ \midrule 
\parbox[t]{3mm}{\multirow{4}{*}{\rotatebox[origin=c]{90}{\makecell{Strongly\\Non-IID}}}} 
& FedAvg \cite{mcmahan2017communication} & 38.44 & 24.88 & 23.89 & 96.85   &&    4.90  & 2.33  & 1.69  & 96.18  && 9.79  & 6.82 & 9.13  & 96.10 \\
& FedProx \cite{li2020federated} & 37.92 & 23.37 & 22.32 & \textbf{96.86}   &&   4.06  & 2.12 & 1.78 & 96.17  && 9.68  & 5.45 & 7.28  & 96.08 \\
& FedSeq \cite{kopparapu2020fedfmc} & 10.73 & 1.05  & 0.59  & 95.73   &&   26.46 & 17.24 & 23.42 & 96.20   && 22.81 & 18.60  & 23.31 & 96.09 \\
\cmidrule{2-16}
& \textbf{UniFed} & \textbf{69.37} & \textbf{55.05} & \textbf{55.87} & 96.75 && \textbf{50.52} & \textbf{37.41} & \textbf{39.18} & \textbf{96.56} && \textbf{46.77} & \textbf{32.45} & \textbf{34.32} & \textbf{96.45} \\
\midrule
\parbox[t]{3mm}{\multirow{4}{*}{\rotatebox[origin=c]{90}{\makecell{Moderately\\Non-IID}}}}
& FedAvg \cite{mcmahan2017communication} & 24.58 & 22.73 & 24.24 & 93.00 && 6.04 & 1.83 & 1.83 & \textbf{94.1} && 10.10 & 5.61 & 6.59 & 92.58 \\
& FedProx \cite{li2020federated} & 24.58 & 22.72 & 24.22 & 92.99   &&   5.10  & 1.31 & 1.05 & 93.80  && 9.37  & 5.34 & 5.55  & 93.46 \\
& FedSeq \cite{kopparapu2020fedfmc} & 3.75 & 0.91 & 0.67 & \textbf{94.67} && 13.75& 7.35 & 8.64 & 92.74 && 19.38 & 19.04 & 21.46 & 93.57 \\
\cmidrule{2-16}
&  \textbf{UniFed} & \textbf{58.02} & \textbf{55.97} & \textbf{58.36} & 94.10 && \textbf{46.15} & \textbf{40.26} & \textbf{43.36} & 93.11 && \textbf{32.40} & \textbf{28.72} & \textbf{31.75} & \textbf{95.3}\\
\bottomrule
\end{tabular}
}
\end{table}

\partitle{FL Environment.} We partition each dataset in a Non-IID fashion across 8 hospitals, resulting in a total of 24 heterogenous hospitals.
All the heterogeneous hospitals are selected to perform FL training. To evaluate the generalizability of \modelname across simple and large deep learning models, following \cite{marfoq2021federated,yang2024fedfed,jia2024fedlps}, we employ CNN (2-layer CNN + 2-layer FFN), VGG11 \cite{simonyan2014very}, and ResNet18 \cite{he2016deep}.
The experiments are performed on Google Colab with a Pro account on a T4 GPU with system RAM 12.7 GB, GPU RAM 15 GB, and Disk 166.8 GB, and each experiment is repeated three times for calculating average evaluation metrics.

\partitle{Evaluation and Comparison Methods.}
In line with state-of-the-art and closely related studies \cite{marfoq2021federated,jia2024fedlps,niu2023flrce}, we benchmark \modelname against Fedavg~\cite{mcmahan2017communication}, FedProx~\cite{li2020federated}, and FedSeq~\cite{kopparapu2020fedfmc} using both strongly and moderately Non-IID data with three FL methods, with $5$ local epochs and $200$ global communication rounds, each evaluated using global model accuracy on test data. As an upper bound, we report the centralized learning method (\ie NoFed) with 200 communication rounds, which can see all tasks and datasets. Our method stands out from conventional FMTL methods such as FedLPS \cite{jia2024fedlps} referenced in related work since we take this challenge to the next level by considering various tasks and datasets \emph{independently}. We set the batch size to 64 and train all hospital models with SGD optimization using a learning rate $\eta =0.001$. We train all local tasks for some local epochs and rounds that \emph{varies} based on task complexity to ensure the timely convergence of all models. Table \ref{tab_perf} shows the evaluation measures including accuracy, f1-score, sensitivity, and specificity of methods trained on MedMNIST datasets with different heterogeneity levels and using different architectures. \modelname consistently outperforms all FL methods in both strongly and moderately Non-IID settings. 
This shows that the \emph{dynamic training} of each hospital for a different number of local epochs based on their task complexity as well as \emph{stopping} at the optimal round is resource-efficient and performance-booster.

\begin{figure}[!th]
  \centering
   \includegraphics[scale=0.3]{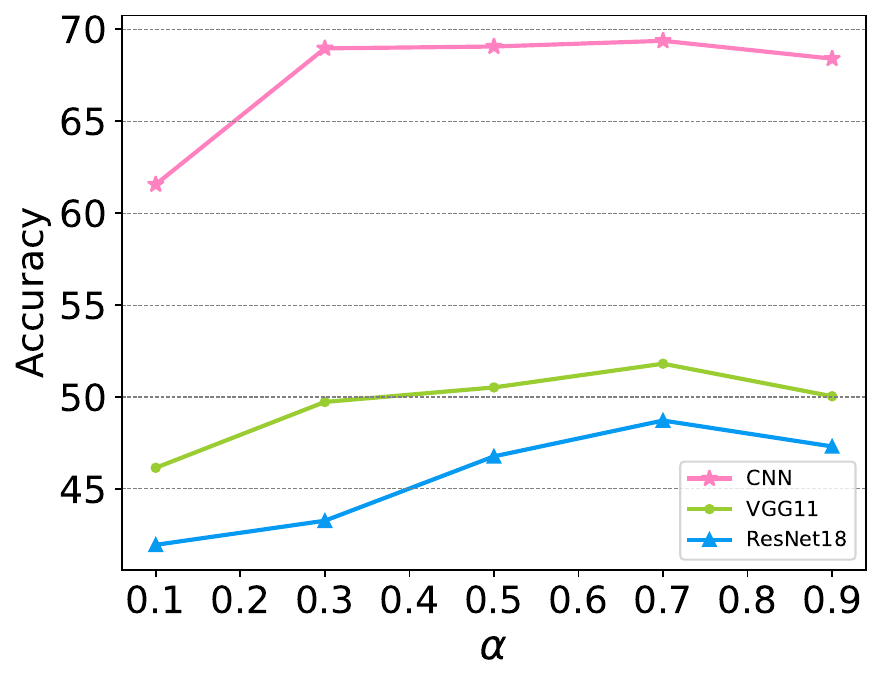}
  \caption{Analysis of various values of $\alpha$ across different models.}
\label{fig:alpha}
\end{figure}

\partitle{Heterogeneity.} To study the effectiveness of \modelname in a mixed setting with different degrees of heterogeneity, we vary the portion of labels across hospitals to investigate the moderately and strongly heterogeneous settings. As shown in Table \ref{tab_perf}, all methods show better performance in the strongly heterogeneous setting. \modelname also achieves higher performance in both moderately and strongly heterogeneous settings across all evaluation measures (except Spec with a minor difference compared to FedProx). \textbf{Computation Overhead.}
We record the cumulative training time for each method and model until converge, as shown in \textbf{Supp.~Table 1}. Our proposed \modelname costs $98.43$ minutes to achieve a great accuracy improvement against benchmarks. Notably, other FL methods have a higher compute overhead. In FL methods that overlook task difficulty and the varying convergence rate of each hospital and server, hospitals may experience overtraining or undertraining, leading to prolonged training time.  \textbf{Communication Overhead.} 
We show the communication overhead for all hospitals over all rounds in \textbf{Supp.~Fig 1} and \textbf{Supp.~Table 1}, respectively. The benchmark FL methods have similar communication costs, however, \modelname achieves the lowest communication cost. 

\partitle{Analysis of Mixing Weight ($\alpha$). }
Furthermore, we evaluate the performance of \modelname by varying the value of mixing weight $\alpha$ = $\{0.1, 0.3, 0.5, 0.7, 0.9\}$ across different models including CNN, VGG11, and ResNet18 shown in Figure~\ref{fig:alpha}.
The mixing weight was fixed at $\alpha=0.7$, giving greater importance to the client's regularization term.

\section{Conclusion}
\label{sec:result}
We presented \modelname, a novel federated learning framework mixing highly heterogeneous medical image classification tasks, dataset, and imaging modalities. Specifically, we proposed a universal sequential dynamic update of local models based on their task complexity. Our method enabled hospitals to dynamically train their local model with \emph{optimal} convergence times. Our extensive experiments demonstrate the effectiveness of \modelname across different architectures against state-of-the-art FL methods on various medical datasets in both strongly and moderately Non-IID settings. Our results also showed the efficiency of \modelname in decreasing both computation and communication overhead. In our future work, we will extend \modelname by leveraging knowledge distillation to avoid sharing a small portion of clients' datasets with the server.
We will also evaluate our \modelname on state-of-the-art foundational models for medical image classification.

%
%
%
\bibliographystyle{splncs}
\bibliography{references}

\end{document}


\title{\modelname: A Universal Federation of a Mixture of Highly Heterogeneous Medical Image Classification Tasks \\ \textit{Supplementary Material}}
\titlerunning{\modelname: A Universal Federation of a Mixture of Heterogeneous Images}
%

\author{Atefe Hassani \and Islem Rekik}
%
\institute{BASIRA Lab, Imperial-X and Department of Computing, Imperial College London, \\ London, UK \\
\email{hasaniatefe0@gmail.com,i.rekik@imperial.ac.uk}\\
}

%
\maketitle   

\section{Communication and Computation Overhead}

\begin{figure}[!ht]
  \centering
   \includegraphics[scale=0.55]{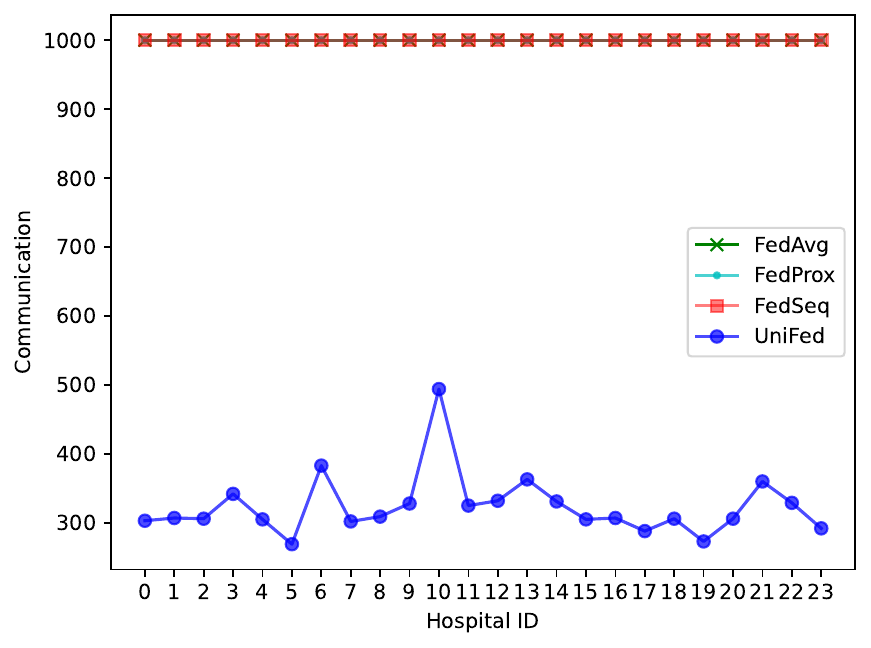}
  \caption{Communication overhead (local epoch for each hospital $\times$ iteration) of \modelname and benchmarks across hospitals.}
\label{fig:com}
\end{figure}

Figure~\ref{fig:com} illustrates the communication overhead across all rounds, highlighting the necessity for dynamic training schedules. Each hospital is trained for varying numbers of global rounds and local epochs based on the complexity of its task, rather than applying a uniform communication schedule for all hospitals. \modelname demonstrates superior performance with significantly reduced communication compared to other methods.

As shown in table~\ref{tab:com}, \modelname achieved significant accuracy gains in just 98.43 minutes, outperforming benchmarks. Unlike other FL methods, it addresses varying task complexities and convergence rates among hospitals and the server, mitigating issues like prolonged training due to overtraining or undertraining.

\begin{table}[ht]
  \centering
    \caption{The computation overhead on MedMNIST dataset with strongly Non-IID setting and the communication overhead (local epoch for each hospital $\times$ number of hospitals $\times$ iteration). The computation overhead shows the total training time in minutes.}
    \begin{tabular}{lccccccc}
    \toprule
    \multirow{2}{*}{Method} & \multicolumn{3}{c}{Computation} && \multicolumn{3}{c}{Communication} \\
    \cmidrule{2-4}\cmidrule{6-8}
    & CNN & VGG11 & ResNet18 && CNN & VGG11 & ResNet18 \\ 
    \midrule
    NoFed & 8.43 & 23.56 & 30.58 && 24000 & 24000 & 24000 \\ \midrule
    FedAvg & 207.65 & 251.56 & 162.02 && 24,000 & 24,000 & 24,000 \\ 
    FedProx & 71.26 & 276.72 & 244.78 && 24,000 & 24,000 & 24,000 \\ 
    FedSeq & 170.58 & 227.2 & 169.03 && 48,000 & 24,000 & 24,000 \\ 
    \midrule
    \modelname & 98.43 & \textbf{50.41} & \textbf{39.78} && 30,327 & \textbf{7,765} & \textbf{6,213} \\ 
    \bottomrule
    \end{tabular}
    \label{tab:com}
\end{table}

\section{Data Analysis}
\begin{figure}
    \centering
    \subfloat[\label{fig:label_distribution}]
    {{\includegraphics[scale=0.44]{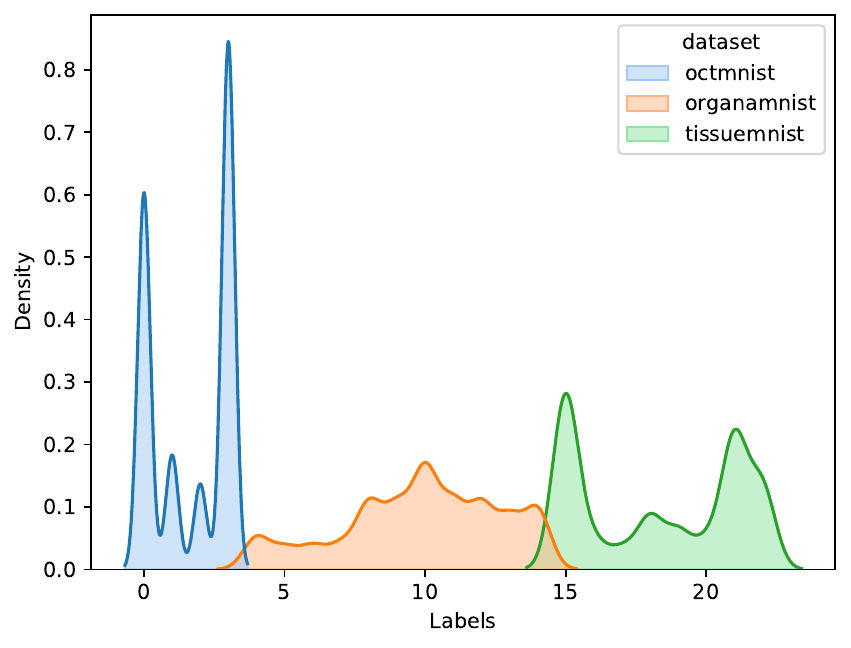} }}%
    \subfloat[\label{fig:pixel_value}]
    {{\includegraphics[scale=0.44]{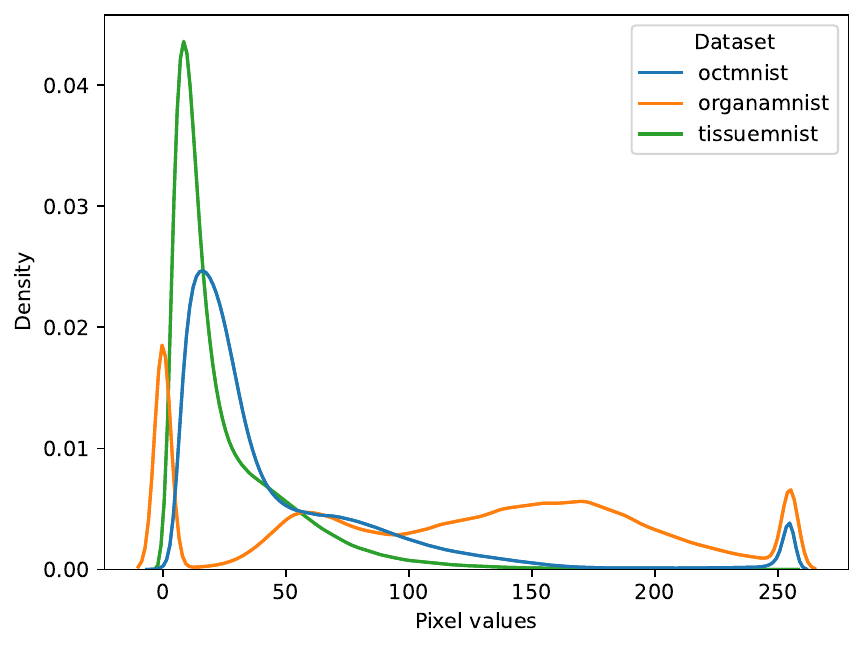} }}%
    \caption{(a) Heterogeneous label distribution across task and (b) Non-IID feature distributions across the 3 datasets.}
    \label{fig:data_distribution}%
\end{figure}

\modelname involves the simulation of various hospitals organized into groups, each group is assigned a distinct classification task. This allocation of tasks leads to a Non-IID class distribution across different tasks, as illustrated in Figure~\ref{fig:label_distribution}. Another aspect of heterogeneity in our setting is the high deviation within the feature space known as feature shift, which we can depict using the image examples from our three benchmark datasets and the pixel value plot available in Figure~\ref{fig:pixel_value}. These two figures demonstrate the shifted distributions and heterogeneity. 

\section{\modelname Algorithm}

\begin{algorithm}[ht]

        \SetAlgoLined
        \KwInput{There are $K$ hospitals indexed by $k$; Local data $D_k$; Server's data $D^S$; Learning rate $\eta$; $E_f$ and $E_{max}$ are the number of local epochs for the first training round and dynamic hospital/server training respectively.}
        \KwOutput{$\theta$; $s$}
        At the server, initialize a global model $\theta_0$ using a random seed SEED, and broadcast it to the hospitals.
        
        \textbf{First Training Round:}
        
        \For{hospital k=1 to K \textcolor{red}{in parallel}}    
            { 
            \For{local epoch from 1 to $E_f$}
                {
                $\theta_0^k$, $s_k \leftarrow$ Hospital Update($\theta_0$, $E_f$)
                }
            }
            
        \textbf{Sequential Training:}
        
        \For{round $t=1$ to $E_{max}$}
            {
            $argsort(\{s_k\}_{k=1}^K)$
            
            \For{hospital $k=1$ to $K$ \textcolor{red}{in sequence}}
                {
                $\theta_t^k \leftarrow$ Dynamic Hospital Update($\theta_t^{k-1}$, $E_{max}$)
                }
            $\theta_t^S \leftarrow$ Dynamic Server Update($\theta_t^S$, $E_{max}$)
            
            $\theta_{t+1} \leftarrow \alpha\theta_t^K + (1 - \alpha)\theta_t^S $
            
            }

        \textbf{Dynamic Hospital/Server Update:}
        
        Initialize Convergence tag = False
        
        \While{tag == False} 
            {
            $B \leftarrow$ Split $D_k$ into batches of size $B$
            
            \For{each batch $b\in B$}
            {
            Compute gradient $\nabla l(\theta; b)$ on batch $b$\\
            Update variable $\theta \leftarrow {\theta} - \eta{\nabla l(\theta; b)} $ \\ 
            Compute model loss $L$ on validation set \\
            }
            \If{Convergence conditions holds}
            {
            tag = true
            }
            }    
        $s, \theta \leftarrow$ Compute loss slope through a list of loss $L$,  $\theta$
        
    \caption{\modelname}
\end{algorithm}